\documentclass{aa}
\usepackage{graphicx}
\usepackage{natbib}
\usepackage{color}

\usepackage{txfonts}
\def\igrj{IGR J$17252-3616$}
\def\nh{N$_{\rm H}$}

\def\xmm{\emph{XMM-Newton}}

\begin{document}
  \title{X-Ray Wind Tomography of the highly absorbed HMXB \igrj \thanks{ 
                 Based on observations obtained with XMM-Newton and INTEGRAL, two ESA science
mission with instruments, data centers, and contributions directly funded by 
                 ESA Member States, NASA, and Russia.}}


  \author{A. Manousakis\inst{1,2} and R. Walter\inst{1,2}
            }

  \institute{ISDC Data Center for Astrophysics, Chemin d'Ecogia 16, CH-1290 Versoix, Switzerland \\
             \email{Antonios.Manousakis@unige.ch}
        \and Observatoire de Gen\`eve, Universit\'e de Gen\`eve,  Chemin des Maillettes 51, CH-1290 Versoix, Switzerland  
            }  

  \offprints{A. Manousakis}

  \date{Received XXX; accepted XXX}


 \abstract
  {About 10 persistently highly absorbed super-giant High-Mass X-ray Binaries (sgHMXB) have been discovered by INTEGRAL as bright hard X-ray sources
lacking bright X-ray counterparts. Besides IGR J16318-4848 that features peculiar characteristics, the other members of this family share many properties 
with the classical wind-fed sgHMXB systems. 
  }
  {Our goal is to understand the specificities of highly absorbed sgHMXB and in particular of the companion stellar wind, thought to be responsible for
 the strong absorption.
  }
  {We have monitored \igrj, a highly absorbed system featuring eclipses, with \emph{XMM-Newton} to study the variability of the column density and of the 
   Fe K$\alpha$ emission line along the orbit and during the eclipses. We also built a 3D model of the structure of the stellar wind to reproduce the
   observed variability.
  }
  {We first derived a refined orbital solution for this system built from \emph{INTEGRAL}, \emph{RXTE} and \emph{XMM-Newton} data. 
  The \emph{XMM-Newton} monitoring campaign revealed 
   significant variation of intrinsic absorbing column density along the orbit and of the Fe K$\alpha$ line equivalent width around the eclipses. 
  The origin of the soft X-ray absorption is modeled with an dense and extended hydrodynamical tail, trailing the neutron star. This structure extends
  along most of the orbit, indicating that the stellar wind is strongly disrupted by the neutron star. The variability of the absorbing column density
  suggests that the wind velocity is smaller ($\upsilon_{\infty}\approx$ 400 km/s) than observed in classical systems. This can also explain the much stronger
  density perturbation inferred from the observations. Most of the Fe K$\alpha$ emission is generated in the most inner region of the hydrodynamical
  tail. This region, that extends over a few accretion radii, is ionized and does not contribute to the soft X-ray absorption.
  }
  {We have built a qualitative model of the stellar wind of \igrj\, that can represent the observations and suggest that highly absorbed systems have a lower
  wind velocity than classical sgHMXB. This proposal could be tested with detailed numerical simulations and high-resolution infrared/optical observations.
  If confirmed, it may turn out that half of the persistent sgHMXB have low stellar wind speeds.
  }

  \keywords{X-Rays:binaries -- Stars: pulsars : individuals: \igrj = EXO 1722-363  --  
               }

\maketitle
%

\section{Introduction}

High mass X-ray binaries (HMXB) consist of a neutron star or a black-hole fueled by the accretion of
the wind of an early type stellar companion.
Their X-ray emission, a measure of the accretion rate, shows a variety of transient to persistent patterns.
Outbursts are observed on timescales from seconds to months and dynamical ranges varying by factors of $10^4$.
The majority of the known HMXB are Be/X-ray binaries \citep{catalogueHMXB}, with Be stellar companions. These systems are transient, 
featuring bright outburst with typical duration on the order of several weeks \citep{XRB_COE}. 
A second class of HMXBs harbour OB supergiants companions (sgHMXBs) feeding the compact object 
through strong, radiatively driven, stellar winds or Roche lobe overflow. 
Thanks to INTEGRAL, the number known sgHMXB systems  tripled during the last decade \citep{Walter_et_al06}.

Highly absorbed sgHMXB were discovered by \emph{INTEGRAL} \citep{Walter_et_al04_atel} and are characterized by strong and persistent 
soft X-ray absorption $(\rm{N_H}>10^{23}~{\rm cm^{-2}})$. When detected, these systems have short orbital periods and long spin 
periods \citep{Walter_et_al06}. They corresponds to the category of  wind-fed accretors in the Corbet diagram\citep{corbetdiagram}. 

\igrj\, has been detected by $ISGRI$ on board $INTEGRAL$ on February 9, 2004 among other hard X-ray sources \citep{Walter_et_al04_atel, Walter_et_al06}. 
The source has first been detected by $EXOSAT$ (EXO 1722-3616) as a weak soft X-ray source, back in 1984 \citep{Warwick_et_al88}. 
In 1987, $Ginga$ performed a pointed observations and revealed a highly variable X-ray source, X1722-363, with a pulsations 
of $\sim 413.9$ sec \citep{Tawara_et_al88}. Further $Ginga$ observations revealed the orbital period of 9-10 days and a
mass of the companion star of $\sim$ 15 $M_{\sun}$ \citep{Takeuchi_et_al90}. Both papers concluded that the system was a high mass X-ray binary (HMXB).

$INTEGRAL$ and $XMM$ observations of \igrj allowed to identify the infrared counterpart of the system and to accurately measure the absorbing column density 
and refine the spin period of the system. Thanks to the eclipses, an accurate orbital period could be derived from $INTEGRAL$ data \citep{Zurita_et_al06}. 
Further RXTE observations revealed a highly inclined system ($i>61^{o}$) with a companion star of $\rm{M}_{*}\la20\, \rm{M}_{\sun}$ and $\rm{R}_{*}\sim 20-40 \,\rm{R}_{\sun}$ \citep{Thompson_et_al07}.
Recent VLT observations provided the companion spectral type \citep{Chaty_et_al08,Mason_et_al09} and radial velocities measurements \citep{Mason_et_al09b}. 
Its spectral energy distribution  can be characterized by a temperature of $\rm{T}_{*}\sim 30$ kK and a reddening of  A$_{V}\sim 20$ \citep{Rahoui_et_al08}.

In this paper, we report on a monitoring campaign of \igrj\, performed with \xmm\, along the orbit in order to estimate the structure of the stellar wind and 
of the absorbing material in the system. We describe the data and their analysis in Section 2,  a refined orbital solution in section 3 and present the evolution 
of the X-ray spectral shape along the orbit in section 4. In section 5, we present and discuss a 3D model of the structure of the stellar wind that can reproduce 
the observations and conclude in section 6.


\section{Data reduction and analysis}

\subsection{\xmm}
Pointed observations of \igrj\,  were performed between August and October 2006 with \xmm\, \citep{xmm-ref}. We scheduled 
9 observations to cover the orbital phases : 
0.01, 0.03, 0.08, 0.15, 0.27, 0.40, 0.65, 0.79, and 0.91. In addition, we used one observation of 2004 with a phase of 0.37. 
The observations are summarized in table \ref{obslogrun}. 

The Science Analysis Software (XMM-SAS)  version 
9.0.0\footnote{http://xmm.esac.esa.int/sas/}  was used  to produce event lists for the EPIC-pn instrument \citep{epic-pn-ref} by running  \verb=epchain=. 
Barycentric correction and good time intervals (GTI) were applied. Photon pile-up and/or out-of-time event were not identified among the data. 
High level products (i.e., spectra and lightcurves) were produced using \verb=evselect=\footnote{http://xmm.esac.esa.int/sas/current/howtousesas.shtml}.
Spectra and lightcurves were built by collecting double and single events in the energy range 0.2 - 10 keV. The lightcurves were built using  5 sec  time bins. The spectra 
were re-binned to obtain 25 counts/bin for low count rate observations and 100 counts/bin for high count rate observation. 

In one dataset (ObsID 0405640701, $\sim 19\,$ ksec), the count rate above 10 keV has a very peculiar behavior, increasing monotonically with time. As this does not 
affect the background subtracted source lightcurve significantly we kept the entire data for pulse arrival times determination. 
Standard GTI was used for spectral analysis.

\subsection{INTEGRAL}
We reanalysed the complete hard X-ray light-curve of \igrj\, obtained with $ISGRI$ \citep{isgri-ref} on board $INTEGRAL$ \citep{integral-ref}.
We extracted the 22--40 ${\rm keV}$ light-curve using the HEAVENS interface \citep{heavens-ref} provided by the $INTEGRAL$ Science Data 
Centre\footnote{http://www.isdc.unige.ch/heavens} \citep{isdc-ref}. 
The lightcurve includes all  public data available on \igrj\, from 2003-01-29 06:00:00 to 2009-04-08 00:28:48 UTC.
The effective exposure time on source is $\sim$ 3.6 Msec.

\begin{table*}
\centering
\caption{XMM-Newton observation log listing the revolution number, the start time, the effective exposure, the source counts, and the phase,  calculated at the middle of each observation using the orbital solution obtained with fixed orbital period (table \ref{table:orbital_solution}).
}
\begin{tabular}{l c  c c c   c c }     
\hline\hline
\noalign {\smallskip}
ObsID      &   \#   &     Revolution & Start time (UT)      &Effective            & Source counts  & Phase                    \\
                 &           &                        &                                  &exposure (ks)    &                            &   ($\pm$ 0.01)                     \\
\noalign {\smallskip}
\hline 
\noalign {\smallskip}
0405640201 & 1 &  1231      & 2006-08-29T03:02:58  &   19.2     & 5.4$\times10^2$ &  0.03 \\
0405640301 & 2 &  1232      & 2006-08-31T16:37:44  &     4.1     & 1.4$\times10^4$ &  0.27             \\ 
0405640401 & 3 &  1234      & 2006-09-04T06:35:33  &     5.6    & 7.9$\times10^3$  &  0.65              \\
0405640501 & 4 &  1235      & 2006-09-06T19:33:10  &     5.4     &9.5$\times10^2$  & 0.91             \\
0405640601 & 5 &  1236      & 2006-09-08T10:03:38  &     7.9     &3.8$\times 10^2$ & 0.08                \\
0405640701 & 6 &  1239      & 2006-09-15T07:23:45  &      2.8   & 1.7$\times 10^3$  & 0.79    \\
0405640801 & 9 &  1247      & 2006-10-01T03:24:26  &      9.4    & 2.3$\times 10^4$  & 0.40                  \\
0405640901 & 8 &  1246      & 2006-09-28T14:36:53 &     11.3    & 1.9$\times 10^4$  & 0.15                   \\
0405641001 & 7 &  1245      & 2006-09-27T07:27:58  &     9.4     & 6.5$\times10^2$ & 0.01   \\
\noalign {\smallskip}
\hline       
\noalign {\smallskip}
0206380401 & 10 & 785        & 2004-03-21T13:23:09 &     8.6   &4.7$\times10^4$  & 0.37        \\
\noalign {\smallskip}
\hline \hline
\end{tabular}
\label{obslogrun}
\end{table*}

\section{Timing analysis and orbit determination}

\subsection{Orbital Period from INTEGRAL} \label{text:orbit_integral}

\begin{figure}[!h]
\centering
  \includegraphics[width= 0.5 \textwidth]{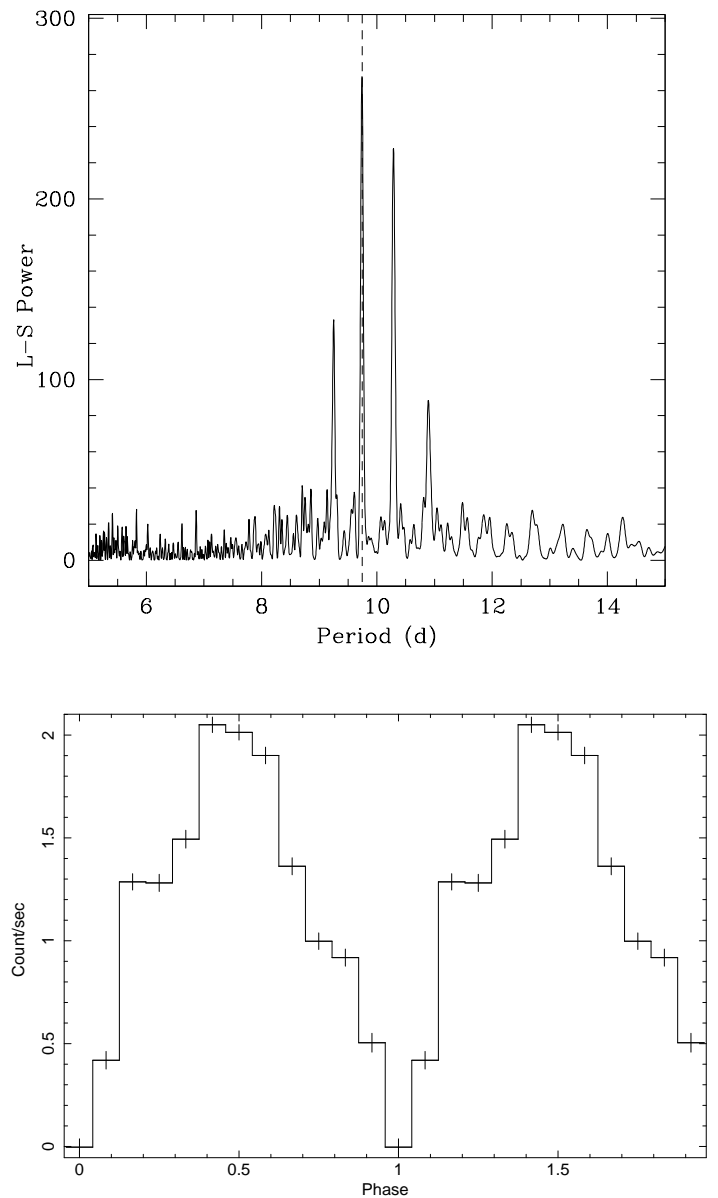}
  \\
   \includegraphics[width= 0.4 \textwidth, angle=270]{fold17252_mine}
    \caption{ {\emph Top}: Lomb-Scargle periodogram obtained from the $INTEGRAL$ 22-40 keV lightcurve. {\emph Bottom:} $INTEGRAL$ lightcurve folded with a
    period of 9.742 days. 
    }
         \label{fig:LS17252}
\end{figure}

We have used the Lomb-Scargle \citep{LS_Press} technique to determine the orbital period from the $INTEGRAL$ light-curve and obtained
P$_{orb}$=$9.742\pm0.001$ days. Figure \ref{fig:LS17252} shows (upper panel) the Lomb-Scargle power around the orbital period (dashed line). 
This period was used to refine the orbital solution (section \ref{text:orbital_solution}).
The lower panel of figure \ref{fig:LS17252} shows the light-curve folded with the newly derived orbital period. The eclipse is clearly detected with the
count rate dropping to zero.

\subsection{Pulse Arrival Times} \label{text:pats}

Pulse arrival times (PATs, hereafter) have been obtained from the broad-band 0.2-10 keV lightcurves obtained by \xmm. 
Close to the eclipse ($\phi=0.03,\, 0.08,\, 0.91,\, 0.01$), when the compact object is behind  the massive star, pulses could not be detected. We  did not 
extract PATs for the observation of 2004.

In order to determine the PATs we used a pulse profile template. This template is derived by folding  the lightcurve from observation 0405640801 with a period of 414.2 sec, 
obtained using the Lomb-Scargle technique \citep{LS_Press}.
This observation was selected  because the source was very bright for a long and almost uninterrupted exposure.

A sequence of pulse profile template was fit to each individual lightcurve. This sequence is characterized by: 
(i) the time of a pulse at the middle of the observation, 
(ii) the pulse period, and  
(iii) the amplitudes of each pulse.
This assumes that the pulse period is reasonably constant during each observation.

The statistical errors on the pulse time and period are typically  0.01 sec and 0.1 sec, respectively.
Table \ref{table:xmmpats} lists the pulse times and the  spin period obtained at the middle of each observation. 
The spin periods were always consistent with the Lomb-Scargle period of each run.

The PAT accuracy is however limited by the systematic error related to the assumed pulse template. Using a different pulse profile template (derived from observation 0405640901) produces  
PAT with offset of 8  to 12 sec from the values listed in table \ref{table:xmmpats}.  We adopted a systematic error of 10 sec.

\begin{table}
\caption{Pulse Arrival Times and derived pulse period. Last column gives the Lomb-Scargle periods. 
We couldn't get a good S/N LS diagram for observation  0405640301.}            
\label{table:xmmpats}                             
\begin{tabular}{ l c c c }    
\hline\hline                      
ObsID                 &           PAT (HJD)             &    Pulse period (s)     &  L-S period    (s)      \\
			&       ( $\pm$ 0.00011 )                         &                                         &                                 \\
\hline
0405640301    &     53978.74948                                   &  414.3 $\pm$ 0.1           &   -                            \\
0405640401    &     53982.32329                                  & 414.0 $\pm$ 0.1             & 417.4 $\pm$0.6   \\
0405640701   &      53993.45320                                  & 414.2 $\pm$ 0.2             &  414.6 $\pm$0.1    \\
0405640801   &      54009.24191                                 & 414.2 $\pm$ 0.1            &  414.2 $\pm$0.4     \\
0405640901   &      54006.75737                                 & 413.8 $\pm$ 0.1               &  413.8 $\pm$0.3     \\
\hline              
\end{tabular} 
\end{table}


\subsection{Orbital Solution} \label{text:orbital_solution}

We have derived the orbital solution using the PATs of the $RXTE$ observation  obtained by  \citet[Epoch 3]{Thompson_et_al07}, and the PATs derived above from $XMM$ data. 

The orbital solution is obtained by comparing the observed pulse arrival delays ($t_{n}- t_{0} - nP_{0} - \frac{1}{2}n^{2}P_{0}\dot{P}$) to the expected ones 
$ \alpha_{x}{\rm sin}i\, {\rm cos} [2\pi(t_{n}-T_{90})/P_{orb}]$ \citep{Levine_et_al_04}. 
The orbital parameters (the orbital period, $P_{orb}$; the projected semi-major axis, $\alpha_{x}{\rm sin}i$; the reference time corresponding to mid-eclipse, $T_{90}$) are assumed to be constant.
To account for pulse evolution, two set of pulse parameters  (spin period at time $t_{0}$, $P_{0}$;  spin period 
derivative, $\dot{P}$) have been used for the $RXTE$ and $XMM$ campaigns.
The pulse number $n$ is given by the nearest integer to  $n=(t_{n}-t_{0})/(P_{0}+0.5\dot{P}(t_{n}-t_{0}))$.

We have  performed a combined fit for the $RXTE$  and $XMM$ observations. 
We derived the orbital solutions by (i) allowing all the parameters to vary freely and (ii) by fixing the orbital period to the value derived from $INTEGRAL$ data. The resulting parameters are listed in table \ref{table:orbital_solution}. 
We could also obtained an upper limit (90\%) of $e<0.15$ on the eccentricity. The latter has been achieved by adding 
the first-order term in a Taylor series expansion in the eccentricity \citep{Levine_et_al_04}. 
The $RXTE$ and $XMM$ orbital solutions are comparable and the resulting parameters are consistent within the errors. 

The folded lightcurve obtained by $INTEGRAL$ derived orbital period (fig. \ref{fig:LS17252}, lower panel) results in a  pulse fraction $\sim$100\%. 
For the rest of the analysis, we used the orbital solution obtained with fixed $P_{orb}$ (table \ref{table:orbital_solution}). 

\begin{figure*}[!ht]
\centering
  \includegraphics[width=17cm]{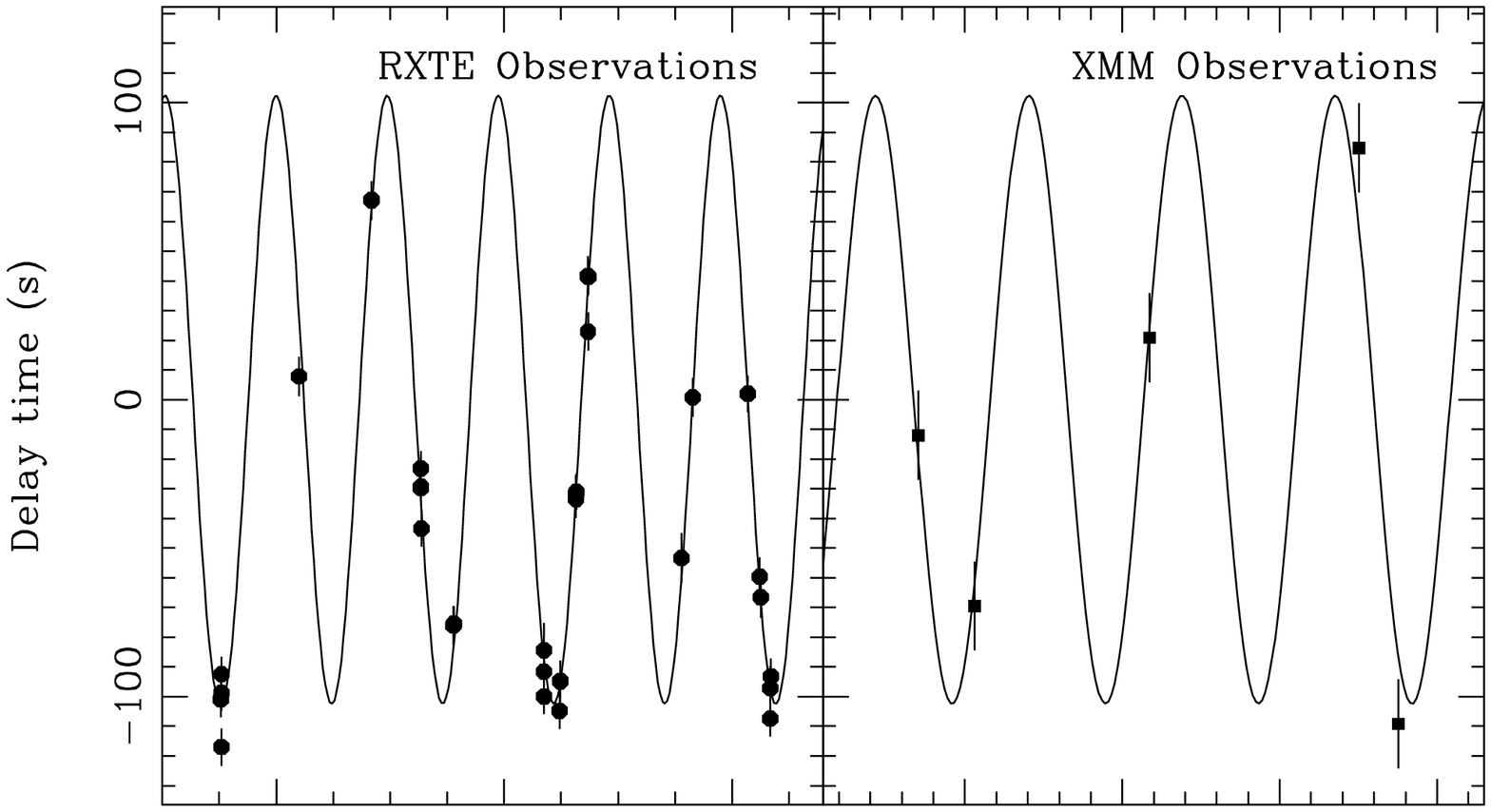}\\
  \includegraphics[width=17cm]{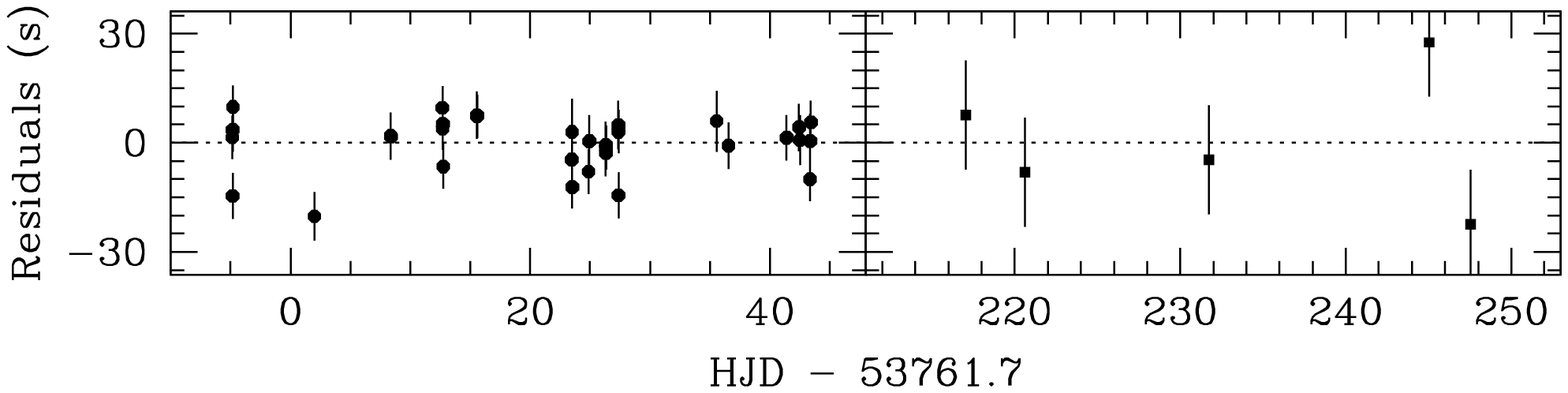}
    \caption{Delays (top) and residuals (bottom) derived from the fixed $P_{orb}$ orbital solution compared to the data. \emph{Left}: $RXTE$ data are 
    taken from \citet{Thompson_et_al07}. \emph{Right}: $XMM-Newton$ data from this work.}
    \label{fig:orbital_solution}
\end{figure*}

Figure \ref{fig:orbital_solution} shows the resulting pulse arrival times delays (fixed $P_{orb}$) for both $RXTE$ and $XMM$ data together with the best-fit orbital solution.

\begin{table}
\caption{Orbital solution for  \igrj.   The errors have been calculated at 90\% confidence level. The errors on the arrival times $t_{0}$ are  $0.00001$  and $0.0001$ days for 
$RXTE$ and $XMM$, respectively. 
The last column shows the result from \citet{Thompson_et_al07} for comparison.
}            
{\tiny
\label{table:orbital_solution}                             
\begin{tabular}{ l l l l l}    
\hline\hline \noalign {\smallskip}        
                           &  Units  &                                         Free    $P_{orb }$                 &    Fixed  $P_{orb}$          &         RXTE Epoch 3 \\    
            	          &            &                                                                                          &                                            &                     \\
\hline \noalign {\smallskip}
${\rm t}_{0}^{XTE}$          &HJD&                                                      53761.73144                    &   53761.73142 &         53761.73126\\
${\rm P}_{0}^{XTE}$     &s &                                                           413.889$\pm$ 0.004           &  413.889$\pm$0.005     &         413.894$\pm$ 0.002\\
$\dot{{\rm P}}^{XTE}$        &  $\mu$s s$^{-1}$                                &           -0.010$\pm$0.002               &  -0.010$\pm$0.003         & -0.0106$\pm$0.0001 \\
\noalign {\smallskip}
\hline \noalign {\smallskip}
${\rm t}_{0}^{XMM}$                            &HJD&                                       53978.7494                         &  53978.7495    &                         -\\
${\rm P}_{0}^{XMM}$                           &s&                                      413.86  $\pm$0.04            &    413.84$\pm$ 0.04       &                 -\\
$\dot{{\rm P}}^{XMM}$        & $\mu$s s$^{-1}$&                      0.98$\pm$ 0.04        &    $1.01\pm 0.03$           &                    -\\
\noalign {\smallskip}
\hline \noalign {\smallskip}
$\alpha_{x}{\rm sin}i$                &lt-s&                                  102$\pm$8                          &   101$\pm$2           &             101$\pm$4\\
${\rm P}_{orb}$                                     &d&                                        9.76 $\pm$ 0.02            &   9.742 (fixed)          &            9.78$\pm$0.04\\
${\rm T}_{90}$	                                    &HJD&                                  53761.62 $\pm$ 0.1         &     53761.69$\pm$0.1          &    53761.60$\pm$0.09\\
\noalign {\smallskip}
\hline \noalign {\smallskip}
$\chi_{\nu}^{2}$                        &&                                  1.5 (30)                                      &            1.65 (31)               &  1.45 (28)     \\
\noalign {\smallskip}
\hline              
\vspace{-0.5 cm}

\end{tabular}  }

\end{table}

\section{Spectral analysis}

The spectral analysis was performed using the XSPEC\footnote{http://xspec.gsfc.nasa.gov} package version 11.3.2ag 
\citep{xspec_ref}. 
In order to use the $\chi_{\nu}^{2}$ statistics, we grouped the data to have at least 25 (faint spectra) and up to 100 (bright spectra) counts per bin.
We initially fitted the observed spectra using a phenomenological model made of an intrinsically absorbed cutoff power law, a blackbody soft excess, and a 
gaussian  Fe K$\alpha$ line (\texttt{wabs*(bb+gauss+vphabs*cutoff)}). 
The centroid of the iron line is compatible with $E_{c} \sim 6.40\pm0.03$ keV at all epochs.

\begin{figure}[t]
\centering
  \includegraphics[width= 0.35\textwidth,angle=270]{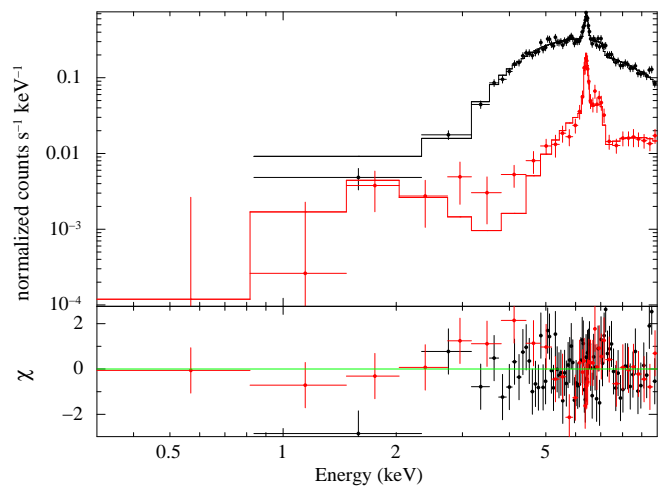}
       \caption{\emph{Top}: Folded model and data for $\phi$=0.65 (black) and $\phi$=0.91 (red). \emph{Bottom}: The corresponding 
       residuals using the model described in the text with the parameters from table \ref{table:spec_results}. }
         \label{fig:spectral_model_folded}
\end{figure}

The spectra are always strongly absorbed below $\sim$ 3 keV and feature an 
iron K-edge at $\sim 7.2\pm0.2$ keV. We first fit each spectrum with all parameters free, 
excepting the galactic absorption, fixed  to  $\rm{N_{H}}=1.5 \times 10^{22}$ cm$^{-2}$ \citep{DLmap}.

Some parameters (photon index, cutoff energy, blackbody temperature, absorber Fe metallicity) did not vary (within the 90\% errors) among the observations and were 
fixed to their average values of $\rm{E_{C}}=8.2$ keV, $\Gamma=0.02$, and $\rm{kT_{BB}}$=0.5 keV, Z=1 Z$_{\odot}$. The Fe  line energy
is fixed to E=6.4 keV, with a narrow width. 
Two representative spectra of our observations are displayed in figure \ref{fig:spectral_model_folded}.

\begin{figure*}[!ht]
\centering
  \includegraphics[width= \textwidth]{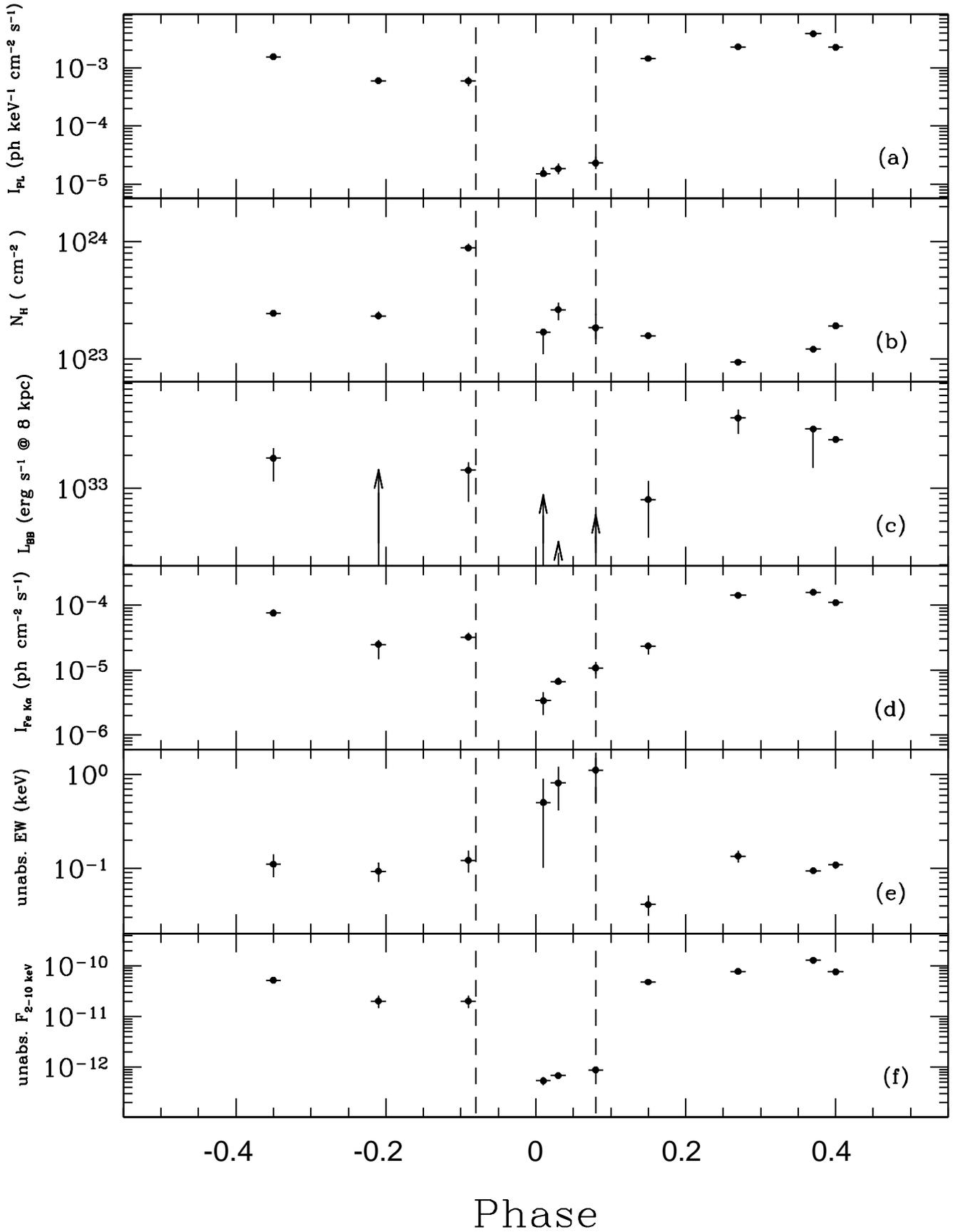}
       \caption{Spectral variability along the orbit. a) The cut-off power-law normalization at 1 keV,  b) The intrinsic hydrogen column density.
        c) The soft excess blackbody luminosity, assuming a distance of 8 kpc . d) The Iron line flux, e) The iron line equivalent width calculated for the unabsorbed continuum,
         and f) the unabsorbed flux. Dashed vertical lines indicates the eclipse boundary.  
        }
         \label{fig:spectral_all}
\end{figure*}

The  intrinsic absorbing column density and the normalization of each component could vary freely.  The best fit parameters are listed in table \ref{table:spec_results}.
Figure \ref{fig:spectral_all} shows the variation of all these parameters and of the unabsorbed  2-10 keV  flux and Fe K$\alpha$ equivalent width (EW). 
The unabsorbed Fe K$\alpha$ EW was calculated by setting the intrinsic  absorbing column density to zero. 
All the spectral fits were good resulting in $\chi_{\nu}^{2} \approx 0.8-1.4$, apart for the $\phi$=0.03 observation 
providing a poor fit. During the eclipse, some  parameters are poorly constrained. 

The unabsorbed continuum flux (fig. \ref{fig:spectral_all}-f) is of the order of  
$\sim 5 \times 10^{-11}$ erg cm$^{-2}$ s$^{-1}$ outside of the eclipse. 
Variations are observed between (2 - 10) $\times 10^{-11}$ erg cm$^{-2}$ s$^{-1}$   and 
they can be interpreted as variation of the accretion rate ($\dot{M}$).
During the eclipse the continuum flux drops by a factor of $\sim200$ and remains stable at the level of 
$F_{unabs}^{2-10 kev} \approx (7\pm1)\times 10^{-13}$ erg cm$^{-2}$ s$^{-1}$.

The intrinsic absorbing column density (fig. \ref{fig:spectral_all}-b) is persistently high ($\ga 10^{23}$ cm$^{-2}$).
Significant variations are detected for  $\phi=0.2-0.4$ and close to the eclipse, reaching values of $N_{H}\sim 9\times 10^{23}$ cm$^{-2}$. 

The normalization of the blackbody component (fig-\ref{fig:spectral_all}-c) does not show any variability, although 
the low energy part of the spectrum is poorly constrained. The normalization of the blackbody component is 
compatible with $\sim 10^{33}$ erg s$^{-1}$ assuming a distance of 8 kpc. 
The very high intrinsic absorbing column density rules out the neutron star as the origin of the soft excess.  

The flux and EW of the Fe K$\alpha$ line are displayed on figure \ref{fig:spectral_all}-d and \ref{fig:spectral_all}-e, respectively.
Both components  show significant variations indicating that the region emitting the line is at least partially obscured by the mass-donor star.  

\begin{table*}
\caption{Spectral analysis results. All the free parameters are listed below together with 
unabsorbed 2-10 keV flux and  $\chi^{2}_{\nu}$. 
All the uncertainties have been calculated at 90\% confidence level. For more details see the text.
}
\label{table:spec_results}
\centering
{\tiny
\begin{tabular}{c|ccccccc}   
\hline\hline
 Phase            & N$_{H} $                                     &     unabs EW (Fe)            &Fe line flux                                        & Blackbody Luminosity                     &   $I _{\nu}^{cutoffpl}$ (1 keV)                                 & $F_{2-10\, keV}^{unabs}$                        &  $\chi_{\nu}^{2}$ (dof)   \\
($\pm$ 0.01)  &$10^{22}$ cm$^{-2}$                &   keV                                  &$10^{-5}$ ph cm$^{-1}$ s$^{-1}$  & $10^{33}$ ergs s$^{-1}$@ 8 kpc  &10$^{-3}$ ph  keV$^{-1}$ cm$^{-2}$ s$^{-1}$&   10$^{-10}$ erg s$^{-1}$ cm$^{-2}$            &   \\
\hline

   0.01                  &17$_{-9}^{+14}$                   &  0.50$\pm$0.4                        & 0.34$\pm$0.21                                  &  $<0.88$                                          & 0.015$_{-0.006}^{+0.009}$        & 0.0054 $\pm$ 0.003  & 1.34 (23) \\
   0.03                  & 26$_{-6}^{+9}$                      &  0.81$\pm$0.4                  &  0.66$_{-0.15}^{+0.12}$                     &  $<0.33$                                              & 0.018$_{-0.005}^{+0.009}$       &   0.006(8)$\pm$ 0.003 &2.6 (17) \\
   0.08                  & 18$_{-9}^{+10}$                    & 1.1 $\pm$0.6                     &  1.1 $\pm$ 0.3                                     &   $<0.57$                                            &   0.023$_{-0.007}^{+0.012}$      &   0.008(8) $\pm$ 0.003    & 1.06 (11) \\
   0.15                  & 15.7 $\pm$ 0.5                      &  0.04 $\pm$0.01                &   2.6$_{-0.6}^{+0.8}$                          &     1.6$\pm$ 1.4                               &  1.44 $\pm$ 0.03                           &   0.48 $\pm$ 0.02        & 0.90 (172)   \\
   0.27                  & 9.4$_{-0.3}^{+0.4}$              & 0.14$\pm$0.02                  &  14$\pm$ 02                                        & 4.4$_{-0.1}^{+0.3}$                         &  2.29$\pm$ 0.05                           & 0.77 $\pm$ 0.02      &   1.05 (128) \\
   0.37                  & 12.1$\pm$0.3                        & 0.09$\pm$0.006                & 16$_{-2}^{+1}$                                   & 3.4$_{-0.3}^{+0.2}$                       & 3.86$_{-0.06}^{+0.05}$                  & 1.29 $\pm$ 0.04       & 1.09 (396) \\
   0.40                  & 19.1$\pm$0.5                        & 0.11$\pm$0.01                   & 11$\pm$1                                           & 2.81$_{-0.1}^{+0.06}$                        &  2.26$\pm$0.05                            &  0.76  $\pm$ 0.02      & 1.45 (207)   \\
   0.65                  & 24$\pm$1                              &  0.11$\pm$0.02                    & 7.6$_{-1.5}^{+0.9}$                         & 1.87$_{-0.06}^{0.1}$                          & 1.55$_{-0.05}^{+0.06}$                &  0.52  $\pm$ 0.02    & 0.99 (73)    \\
   0.79                  & 23$\pm$3                              &  0.09$\pm$0.03                   &  2.5$_{-0.7}^{+1.0}$                           & $<1.48$                                           & 0.60 $\pm$0.06                             &  0.20 $\pm$ 0.04    & 0.97 (31) \\
   0.91                  &89$_{-14}^{+11}$                 &  0.12$\pm$0.03                     & 3.2$_{-0.7}^{+0.5}$                            & 1.41$_{-0.05}^{+0.1}$                      & 0.6$_{-0.1}^{+0.2}$                      & 0.20 $\pm$0.07     & 0.82 (33)   \\
\noalign {\smallskip}
\hline \hline  

\hline
\end{tabular}  }
\end{table*}

As the X-ray continuum illuminating the gas emitting the Fe fluorescent line cannot be measured during the eclipse we calculated 
a corrected Fe K$\alpha$ EW by assuming a constant continuum flux of $1.8\times 10^{-3}$ ph\, keV$^{-1}$ cm$^{-2}$ s$^{-1}$
(fig. \ref{fig:corrected_ew}).

\section{Discussion}

\subsection{Constraining the physical parameters of the system.} 

In the previous sections we have derived 
an orbital solution (table \ref{table:orbital_solution}) yielding a mass function
$f=4\pi^{2} (\alpha_{x} sin\, i)^3/GP_{orb}^2 = M_{OB}\, sin^{3}\,i / (1+q)^2 \approx 11.7\pm 0.7$ M$_{\odot}$. This is consistent with  a high mass x-ray binary system.

Adopting an inclination $i=90^{o}$ we infer a mass $M_{OB}\sim 14$ M$_{\odot}$ for the donor star.
Radial velocity observation showed  $q=M_{X}/M_{OB}\sim 0.1$ \citep{Mason_et_al09b}. 
Using an upper limit on the mass of $\sim 20 M_{\odot}$  \citep{Thompson_et_al07} constrain the inclination of the orbit $i>70^{\circ}$.
\citet{Mason_et_al09b} estimated an inclination $i\approx 75^\circ-90^\circ$. With this value the mass of the donor star is constrained 
in the range  $M_{OB}\approx 14-17$ M$_{\odot}$. 
The mass ratio and the donor mass imply a neutron star mass  $M_{NS}=1.4-1.7$ M$_{\odot}$.
Both the masses of the donor star and of the compact object are roughly similar (within a factor of 2)  to Vela X-1 \citep{Quaintrell_et_al03,vanKerkwijk_et_al95}. 

The separation of the system could be derived from the duration of the eclipse and the inclination \citep{JossRappaportARAA84}.
For the eclipse duration of $\Delta\phi\approx 0.18 \pm 0.02$ (i.e. 1.75 days) we can estimate the separation $\alpha_{x} \approx 1.7-1.8\, R_{*}$.
The Roche lobe ($R_{L}$) of the system scales from $R_{L}=0.99-1.06$ R$_{*}$ assuming a synchronous rotation \citep{JossRappaportARAA84}. 
This means that the system is very close to fill its Roche lobe and form an accretion disk, although no significant spin-up has been observed. 

Using our VLT observations, \citet{Mason_et_al09} performed near-IR spectroscopy of \igrj\, and concluded that the donor star is a B0-B5 I to B0-B1 Ia 
with effective temperature $T_{eff}=22-28$ kT and a stellar radius of $R_{*}=22-36$ R$_{\odot}$.
Based on a spectroscopic determination, we can translate it into absolute numbers for the separation ($\alpha_{x} \approx 37-64$ R$_{\odot}$) and 
the Roche lobe radius ($R_{L}\approx 22-38$ R$_{\odot}$). 

The unabsorbed 2-10 keV source flux is in the range  $(0.2-1.3)\times 10^{-10}$ erg s$^{-1}$ cm$^{-2}$.
Adopting a mean value of ($0.8\pm0.3)\times10^{10}$  erg s$^{-1}$ cm$^{-2}$
and assuming a distance of 8 kpc  \citep{Mason_et_al09} the inferred 2-10 keV unabsorbed luminosity is  L$_{X}\sim  10^{36}$ erg s$^{-1}$.
Assuming accretion as the source of energy ($L_{X}=\epsilon \dot{M} c^{2}$, 
we can estimate a mass accretion rate of $\dot{M}\sim 10^{-9}$ M$_{\odot}$ yr$^{-1}$ similar to that of Vela X-1 \citep{Furstetal10}.

During the eclipse, the X-ray luminosity drops by a factor of $\sim$ 200 resulting a  ${\rm L_{X}}\sim 5\times 10^{33}$ erg s$^{-1}$. 
As OB stars are emitting in X-rays with a luminosity ${\rm L_{X}}\sim 10^{31-32}$ erg s$^{-1}$ \citep{GudelnNaze09}, this emission is probably dominated 
by scattering in the stellar wind \citep{haberl91}.

\citet{softexcess} have discussed the origin of soft X-ray excesses in many types of accreting pulsars. 
It is likely, that the  fairly constant soft X-ray excess observed in \igrj\, is emitted by recombination lines \citep{schulz} 
in a region of the wind larger than the stellar radius \citep{watanabe}.

For a spherically symmetric stellar wind, one would expect to have smooth and predictable variations
 on the absorbing column density along the orbit. In particular observations on the same line of sight or
  symmetric when compared to the eclipse must result in different, respectively identical, column densities. 
  Our observing strategy resulted in \nh($\phi$ = 0.15) $\approx$ \nh($\phi$ = 0.37), \nh($\phi$ = -0.35)$ >$ \nh($\phi$ = 0.37) 
  and \nh($\phi$ = -0.35)$\approx$ \nh($\phi$ = -0.21), ruling out a spherical geometry.

\subsection{Stellar wind structure}

We have constructed  a 3D model of the OB supergiant stellar wind in order to 
simulate  the variability of the intrinsic column density (\nh) and of the 
Fe K$\alpha$ line equivalent width (EW), along the orbit, and to compare them with the observations.
We have assumed a distance $D\approx$ 8 kpc \citep{Mason_et_al09}, a circular orbit ($e=0$) and an edge-on geometry ($i=90^\circ$).

\subsubsection*{Variability of \nh\, along the orbit}   \label{text:nhprofile}

We first investigate the behavior of the intrinsic column density, \nh, as a function of phase, to reveal the structure of the wind along the orbit.  
We approximated the wind structure with two components, the unperturbed wind ($\rho_{wind}$) and a tail-like hydrodynamic perturbation ($\rho_{tail}$) 
related to the presence of the neutron star. Tail-like structures are predicted by hydrodynamical  simulations \citep{Blondin90} but produce \nh, up to 
$\sim$10$^{22}$ cm$^{-2}$, too small to account for the variability observed in IGR J17252--3616.  

The unperturbed stellar wind has been modeled  by assuming a
standard  wind profile \citep{CAKwind}, 
\begin{displaymath}
\upsilon(r)=\upsilon_{\infty}\Big(1-\frac{R_{*}}{r}\Big)^{\beta}
\end{displaymath}
where $\upsilon(r)$ is the wind velocity at distance $r$ from the stellar center, 
$\upsilon_{\infty}$ is the terminal velocity of the wind, and $\beta$ is a parameter describing the wind gradient. 
The conservation of mass provides the radial density distribution of the stellar wind.
The unperturbed stellar wind is a good approximation within the orbit of the neutron star.
Hydrodynamical simulations \citep{Blondin90, Blondin91,blondin94,blondin95, mauche08} of HMXB  have shown that the wind can be highly disrupted by the neutron star
beyond the orbit.

\begin{figure}[!h]
\centering
\includegraphics[height=0.38\textwidth,width=0.475\textwidth,keepaspectratio=false]{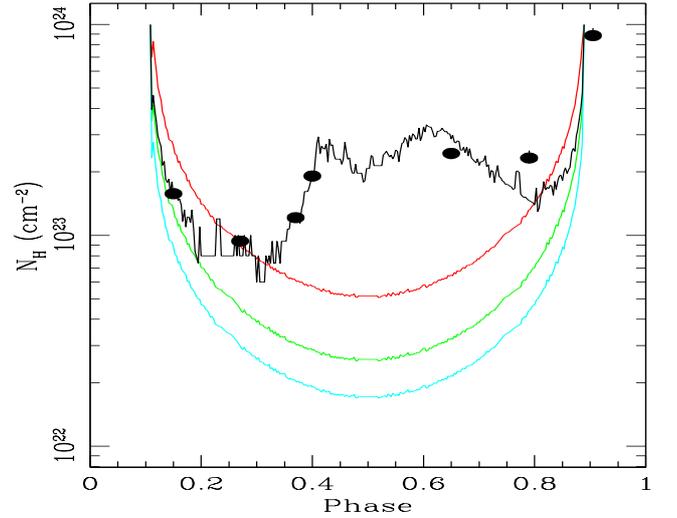}
  \caption{Simulated N$_{H}$ variations plotted together with the data.   
  We illustrate three different wind configurations, for $\dot{M}/\upsilon_{\infty}=0.7$ (cyan), 1 (green), and 
  $2$ (red) $\times 10^{-16}$ M$_{\odot}$/km, keeping all the other parameters fixed. 
  The solid black line shows the total \nh\, consisting of the unperturbed stellar wind (green line) and of the 
  tail-like extended component. The observations during the eclipse have been omitted. 
  }
       \label{fig:nhvsphiforsmoothwind}
\end{figure}

To estimate the terminal velocity of the unperturbed wind, 
we have studied the \nh\ variability using three different sets of parameters (Fig. \ref{fig:nhvsphiforsmoothwind}).
The mass-loss rate and terminal velocity are constrained by the data to be in the range
$\dot{M}_{w}/\upsilon_{\infty}\sim (0.7-2)\times10^{-16}$ M$_{\odot}$/ km; ($\beta$ has a very limited impact on the results, we used 0.7).

The fraction of the wind captured by the neutron star can be estimated from the accretion radius $r_{acc}= 2 GM_{X}/(\upsilon_{orb}^2+\upsilon^2)\sim 2\cdot 10^{11} ~{\rm cm}$ 
(where $\upsilon_{orb}=250~{\rm km/s}$ is the orbital velocity) as $f \sim \pi r_{acc}^{2} / 4\pi R_{orb}^{2} \sim 7.5\cdot 10^{-4}$. 
The mass-loss rate is therefore $\dot{M}_{w}\sim f^{-1}\dot{M}\approx 1.5\cdot10^{-6}$ M$_{\odot}$/yr and the terminal velocity of the wind
constrained in the interval $\upsilon_{\infty}\sim$ 250 -- 600 km/h. 

In our simulation we have adopted a terminal velocity $\upsilon_{\infty}$=400 km/sec, a stellar radius $R_{*}$=29 $R_{\odot}$, 
a  wind gradient $\beta$=0.7, and a mass loss rate $\dot{M_{*}}$= 1.35$\times10^{-6}$ M$_{\odot}\,$ yr$^{-1}$. 
A summary of the assumed, observed, and inferred parameters of the model is listed in table \ref{tab:adhocmodel}.
The variability of \nh\, along the orbit indicates that the unperturbed wind is adequate for phases $\phi\approx 0\, -\, 0.35$.
For $\phi \ga$ 0.35, \nh\, is increasing by   $\sim 2\times10^{23}$ cm$^{-2}$. This indicates that a high density tail-like structure 
lies on one side of the orbit,  trailing the neutron star.
The tail-like component is still present up to $\phi \ga 0.8$.

\begin{figure}
\centering
\includegraphics[height=0.38\textwidth,width= 0.36\textwidth,angle=270]{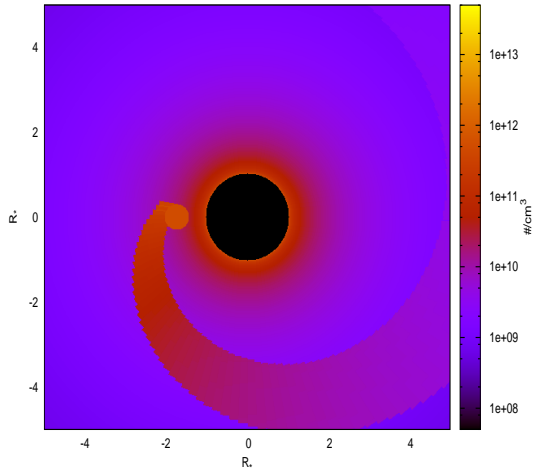}
 \caption{Number density distribution in the plane of the orbit including a smooth wind and a tail-like perturbation. 
The black disk at the center represents the supergiant star. 
}
 \label{fig:accretion_geometry}
\end{figure}

We have assumed that, 
the tail-like structure is created very close to the NS and is opening with distance from the neutron star.
The density of the material inside the `tail' is decreasing with distance for  mass conservation.  
Its distribution follows a `horn'-like shape with a  circular section.
We have adjusted the density of the tail-like structure to match the observations.
The density distribution $\rho_{wind}+\rho_{tail}$ is displayed in
figure \ref{fig:accretion_geometry}. 
The supergiant is located at the center (black disk). The tail-like structure covers about half  of the orbit.

Figure \ref{fig:nhvsphiforsmoothwind} displays 
the simulated \nh variability from the above density distribution together with the observed data points. 
The data and the model shows that the tail-like perturbation is essential to understand the observed variations of the \nh.

\subsubsection*{Variability of the Fe K$\alpha$ line along the orbit} \label{text:feprofile}


Assuming that the intrinsic X-ray flux is not affected by the eclipse, the Fe K$\alpha$ equivalent width drops by 
a factor $\sim 10$ during the eclipse in an orbital phase interval of $\sim 0.1$. This
indicates that the radius of the region emitting Fe K$\alpha$ is smaller than half of the stellar radius $(<10^{12}~{\rm cm})$ 
and much more compact than the tail structure responsible for the ${\rm N_H}$ variability profile. 

Vela X-1 shows a similar behavior that was interpreted with an emitting region of the size of \citep{ohashi84},
or even within \citep{Endo2002}, the accretion radius.

Outside of the eclipse, the equivalent width of the Fe K$\alpha$ line is of the order of 100 eV.
Following \citet{Matt2002} and assuming a spherical transmission geometry, this corresponds to a column 
density of N$_{\rm H}\sim 2\cdot 10^{23}$ cm$^{-2}$. As this additional absorption is not observed,
the region emitting Fe K$\alpha$ is partially ionized. 

A ionization parameter $\xi={\rm L/n R^2}$ in the range 10--300 is required to fully ionize light elements contributing 
to the soft X-ray absorption and to keep an Fe K$\alpha$ line at the energy of 6.4 keV \citep{Kallmanetal04,Kallman82}.
The density of the Fe K$\alpha$ emitting region is therefore $\sim \xi^{-1} \rm{R_{12}}^{-2}\cdot 10^{12}\, {\rm cm}^{-3}$, 
where ${\rm R_{12}}$ is the distance from the neutron star in units of 10$^{12}$ cm. 
As $\rm{N_H=nR\sim2\times 10^{23} cm^{-2}}$ we have $\xi \sim  5/{\rm R}_{12}$. A dense cocoon is therefore needed
around the neutron star with a size $0.5>{\rm R}_{12}>0.02$ and a density $10^{11}\, {\rm cm^{-3}} <{\rm n}<3\cdot 10^{12} \,{\rm cm^{-3}}$. 

To match the observed column density variability, the radius and the density of the inner region of the tail structure were 
set to $4\cdot10^{11}$ cm and $3\cdot10^{11}$ cm$^{-3}$ respectively. It is therefore very likely that the dense 
cocoon corresponds to the inner and ionized region of the hydrodynamical tail.

\begin{figure}
\centering
\includegraphics[height=0.38\textwidth,width= 0.36\textwidth,angle=270]{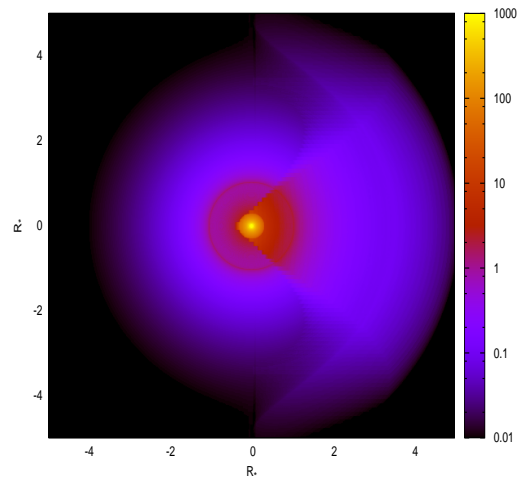}
\caption{Integrated Fe K$\alpha$ emissivity centerred on the neutron star at phase $\phi=0.5$ (relative units). 
The smooth circular halo shows the rim of the supergiant star. The tail structure can be observed on the right.}
 \label{fig:inter_shape_fe_emit}

\end{figure}

We have thus added this partially ionized cocoon in our simulations, using a density
of $3\cdot 10^{11}~{\rm cm}^{-3}$ within a radius of $\sim 6\cdot10^{11}$ cm $\sim 3 \,{\rm R_{acc}}$.
The Fe K$\alpha$ emissivity map (Fig. \ref{fig:inter_shape_fe_emit}) is built applying an illuminating radiation field $(\sim 1/r^{2})$
to the density distribution.

\begin{figure}[!h]
\centering
\includegraphics[height=0.38\textwidth,width=0.475\textwidth,keepaspectratio=false]{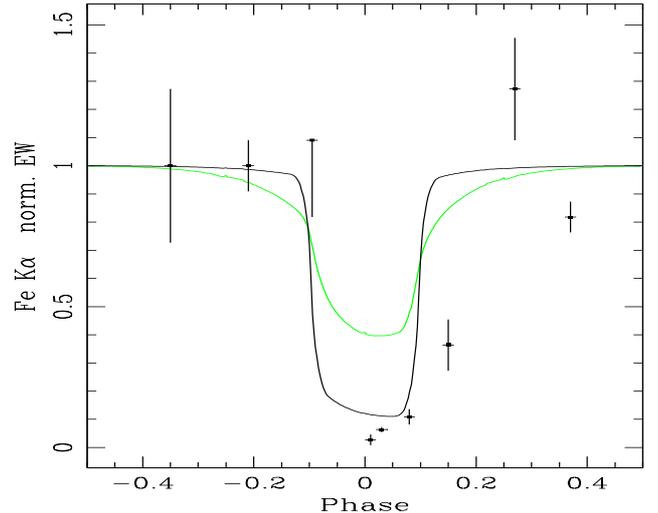}
  \caption{ Corrected unabsorbed Fe K$\alpha$ line equivalent width along the orbit. 
 The two curves represent two different configurations in the presence of the hydrodynamical structure. 
 \emph{Black curve:} the  central cocoon is present.
 \emph{Green curve:} without the central cocoon. See the text for more details. 
  }
       \label{fig:corrected_ew}
\end{figure}

Figure \ref{fig:corrected_ew} displays the resulting simulated profile of the Fe K$\alpha$ equivalent width
together with the observed data. 
The green curve shows the variations of the Fe K$\alpha$ equivalent width expected from the wind density profile 
excluding the central cocoon, which
obviously could not reproduce the data.
The black curve accounts for the dense central cocoon. 
The exact profile of the eclipse is related to the size and density profile of the cocoon. 
No effort has been made to obtain an exact match to the data.

The ionized cocoon is expected to produce an iron K-edge at $\sim 7.8$ keV. 
For a column density of \nh$\sim 2\times 10^{23}$ cm$^{-1}$, the optical depth of this edge $\tau\sim 0.2$ \citep{Kallmanetal04}
remains difficult to detect. Even our observation at an orbital phase 0.14 (the best candidate for the detection of the ionised edge) 
does not have enough signal.

The mass of the tail-like structure M$_{tail}\sim 10^{-8} \, {\rm M}_{\odot}$
can be accumulated in $t_{tail}=M_{tail}/\dot{M}$, 
where $\dot{M}=(\pi r_{eff}^{2}/4\pi R_{orb}^{2}) \dot{M}_{w}$, where 
$r_{eff}$ is an effective radius for the funneling of the wind in the tail.
For a tail accumulation time scale (t$_{tail}$) comparable to the orbital period of ($\sim 10$ days),
this effective radius becomes 12 R$_{acc}$.

The orientation of the tail-like structure depends on the wind and orbital velocity.   
The angle between the wind velocity and the orbital velocity is given by $tan(\alpha)=\upsilon_{orb}/\upsilon$. 
The tail obtained in our simulations is tilted by $\alpha \sim 80^{o}$. This would correspond to $\upsilon\sim 0.2~\upsilon_{orb}\approx 50~{\rm km/s}$ 
which is lower than $\upsilon (R_{orb})\approx 250~{\rm km/s}$ because of the ionization of the stellar wind in the vicinity of the neutron star.

\begin{table}
\caption{Summary of the wind model used for \igrj. 
}             
\label{tab:adhocmodel}      
\centering                          
\begin{tabular}{l l l}        
\hline
\hline                 
Parameter 			& Value 		& Reference  \\    
\hline        
\emph{Observed}\\                
\hline
$q$=M$_{X}$/M$_{OB}$     & 0.1                &\citet{Mason_et_al09b}\\
M$_{OB}$ (M$_{\odot}$)      & 15 		& \citet{Takeuchi_et_al90}   \\
$\alpha$ (R$_{*}$)                    &  1.75                 & This work.\\
$\dot{M}$ (M$_{\odot}$/yr)       &$10^{-9}$& This work. \\
\hline        
\emph{Assumed}\\                
\hline
R$_{*}$ (R$_{\odot}$)      &  29                   &    \\
M$_{X}$ (M$_{\odot}$)          &1.5                    &     \\
$i$ (deg)                                    &   90                 &    \\
$\beta$				& 0.7			& \\
\hline
\emph{Inferred} \\
\hline
$\upsilon_{\infty}$ (km/s)  & 400 		& \\
$\dot{M_{w}}$ (M$_{\odot}$/yr)		&$1.35\times10^{-6}$& \\
\hline                                   
\end{tabular}
\end{table}

\section{Conclusions}

We have presented the analysis of an observing campaign performed with $XMM-Newton$
on the persistently absorbed sgHMXB \igrj. Nine observations have been performed over 
about four weeks, distributed on various orbital phases. Three of them were scheduled
during the eclipse of the neutron star by the companion star. 

We first refined the orbital solution, using in addition archival $INTEGRAL$ and $RXTE$ data 
and found an orbital period of 9.742 d and a projected orbital radius of $101\pm2$ lt-s. 
The pulsar spin period varies between 414.3 and 413.8 sec during the observing campaign.

The X-ray spectrum (0.2 -- 10 keV), which varies along the orbit, was successfully fit using an 
absorbed cut-off power law continuum, a soft excess and a gaussian emission line. The soft 
excess, modeled with a black body, remained constant. 

The continuum component varies in intensity (a measure of the instantaneous accretion rate) 
but features a constant spectral shape, as usually observed in accreting pulsars.

The absorbing column density and the Fe K$\alpha$ emission line show remarkable variations.
The column density , always above $10^{23}$ cm$^{-2}$, increases towards $10^{24}$ close to 
the eclipse, as expected for a spherically symmetric wind. The wind velocity is unusually small
with $\upsilon_{\infty}=400$ km/s. An additional excess of absorption of $2\cdot10^{23}$ cm$^{-2}$ is 
observed for orbital phases $\phi>0.3$ revealing an hydrodynamical tail, trailing the neutron star.

During the eclipse, the equivalent width of the Fe K$\alpha$ line drops by a factor $>10$ indicating
that most of the line is emitted in a cocoon surrounding the pulsar, with a size of a few accretion 
radia. This cocoon is ionized and corresponds to the inner region of the hydrodynamical tail

The parameters of the \igrj\, are very similar to these of Vela X-1, excepting for the smaller
wind velocity. We argue that the persistently large absorption column density is related to the 
hydrodynamical tail, strengthened by the low wind velocity. The tail is a persistent structure 
dissolving on a time scale comparable to the orbital period.

Our interpretation can be tested using numerical hydrodynamical simulation and high resolutions 
optical/infrared spectroscopy. If confirmed, it may turn out that half of the persistent sgHMXB have 
stellar wind speeds several times lower than usually measured.

\begin{acknowledgements}
This research has made use of NASA's Astrophysics Data System.

\end{acknowledgements}
\bibliographystyle{aa} 
\bibliography{paper} 
\end{document}